\documentclass[aps, prd, preprint, groupedaddress, nofootinbib]{revtex4-1}

\usepackage{amsmath}
\usepackage{amssymb}
\usepackage{stmaryrd}
\usepackage{wasysym}
\usepackage{amsfonts}
\usepackage{graphicx}
\usepackage{array}
\usepackage{multirow}
\usepackage{xcolor}

\begin{document}



\title{Relativistic MOND Theory from Modified Entropic Gravity}






\author{
A. Rostami$^{1}$\footnote{abasat.rostami@aut.ac.ir},
K. Rezazadeh$^{2}$\footnote{kazem.rezazadeh@ipm.ir},
and M. Rostampour$^{1}$\footnote{rostampourmostafa@gmail.com}
}

\affiliation{
$^{1}$\small{Physics and Energy Engineering Department, Amirkabir University of Technology (Tehran Polytechnics), P.O. Box 159163-4311, Tehran, Iran}\\
$^{2}$\small{School of Astronomy, Institute for Research in Fundamental Sciences (IPM), P.O. Box 19395-5531, Tehran, Iran}
}

\date{\today}


\begin{abstract}

We derive a relativistic extension of Modified Newtonian Dynamics (MOND) within the framework of entropic gravity by introducing temperature-dependent corrections to the equipartition law on a holographic screen. Starting from a general modification of the surface degrees of freedom and employing the Unruh relation between acceleration and temperature, we obtain modified Einstein equations in which the geometric sector acquires explicit thermal corrections. Solving these equations for a static, spherically symmetric spacetime in the weak-field, low-temperature regime yields a corrected metric that smoothly approaches Minkowski space at large radii and naturally contains a characteristic acceleration scale. In the very-low-acceleration regime, the model reproduces MOND-like deviations from Newtonian dynamics while providing a relativistic underpinning for that phenomenology. We confront the theory with rotation-curve data for NGC~3198 and perform a Bayesian parameter inference, comparing our relativistic MOND (RMOND) model with both a baryons-only Newtonian dynamics (ND) model and a halo of dark matter (DM) model. We find that RMOND and DM models both fit the data significantly better than the baryons-only ND prediction, and that RMOND provides particularly improved agreement at $r\gtrsim 20\,\mathrm{kpc}$. These results suggest that temperature-corrected entropic gravity provides a viable relativistic framework for MOND phenomenology, motivating further observational tests, including gravitational lensing and extended galaxy samples.

\end{abstract}



\maketitle



\section{Introduction}
\label{sec:int}

The persistent mismatch between Newtonian gravity and astrophysical observations, most notably the inability of Newtonian dynamics to reproduce galactic rotation curves, has long motivated the search for alternative explanations. Among these, the dark matter (DM) hypothesis postulates the existence of a non-luminous component that reconciles theory with observation. In contrast, the Modified Newtonian Dynamics (MOND) paradigm, proposed by Milgrom~\cite{Milgrom:1983ca, Milgrom:1983pn, Milgrom:1983zz}, postulates that Newtonian dynamics must be modified at accelerations below a certain threshold value. Although MOND successfully accounts for galactic rotation curves and related empirical relations, it faces serious challenges in explaining gravitational dynamics on larger scales, such as those of galaxy clusters and the cosmic microwave background (CMB)~\cite{Sanders:2002pf, Clowe:2006eq, Angus:2006qy, Skordis:2020eui}.

Over the past few decades, a growing body of research has uncovered profound links between thermodynamics and gravity, reshaping our understanding of spacetime at a fundamental level. The pioneering works of Bekenstein and Hawking demonstrated that black holes behave as thermodynamic systems: they possess entropy and emit thermal radiation~\cite{Bardeen:1973gs, Hawking:1974rv, Hawking:1975vcx, Bekenstein:1973ur}. These achievements indicate that gravitational systems inherently carry thermodynamic properties, suggesting that gravity may have a statistical origin.

Another cornerstone of thermodynamic gravity is the Unruh effect~\cite{Unruh:1976db}, which associates temperature with acceleration. This effect implies that an observer undergoing constant acceleration in the quantum vacuum perceives a thermal bath at a temperature which is proportional to the observer's acceleration~\cite{Unruh:1976db}. This remarkable relation unites acceleration, quantum theory, and thermodynamics, and it plays a central role in modern attempts to view gravity as an emergent phenomenon. It also provides a natural context for modifying gravity through background temperature effects.

A major conceptual advance came with Jacobson’s observation~\cite{Jacobson:1995ab} that the Einstein field equations themselves can be interpreted as equations of state describing a thermodynamic system. This insight paved the way for Verlinde’s entropic gravity framework~\cite{Verlinde:2010hp}, in which the gravitational interaction emerges as an entropic force arising from changes in information associated with the positions of material bodies. By invoking the holographic principle and the equipartition of energy on a holographic screen, Verlinde succeeded in reproducing both Newtonian gravity and Einstein’s general relativity from purely thermodynamic considerations~\cite{Verlinde:2010hp}.

The holographic principle, originally proposed by 't~Hooft and Susskind~\cite{tHooft:1993dmi, Susskind:1994vu, Susskind:2005js}, states that all physical information contained within a spatial region can be fully represented by degrees of freedom residing on its boundary surface. This revolutionary concept implies that the three-dimensional universe can be viewed as a holographic projection of data encoded on a two-dimensional surface, profoundly changing our understanding of space, time, and information.

In Verlinde’s approach~\cite{Verlinde:2010hp}, the total energy transferred to the holographic screen is assumed to be equally distributed among its microscopic degrees of freedom. In the present work, we adopt this viewpoint but allow for temperature-dependent corrections. Studies of the statistical mechanics of condensed matter have revealed that it is necessary to consider corrections to the equipartition law of energy to justify their heat capacity based on the statistical behavior of bosons at low temperatures~\cite{Reif1965, Pathria:1996hda}. In particular, the temperature-dependent corrections to the equipartition law exhibit significant effects at low temperatures.

Moreover, in Verlinde's emergent gravity framework~\cite{Verlinde:2016toy}, the presence of positive dark energy gives rise to a thermal \emph{volume law} contribution to the entropy, which becomes dominant precisely at the cosmological horizon. This contribution competes with the usual \emph{area law} entanglement entropy, and the interplay between the two leads to nontrivial gravitational phenomena at sub-Hubble scales. In particular, due to this competition, the microscopic de~Sitter states fail to thermalize below the Hubble scale and instead exhibit memory effects, manifesting as an entropy displacement induced by the presence of matter. The emergent gravitational dynamics then include an additional, ``dark'' elastic response force associated with this entropy displacement. Verlinde shows that the magnitude of this extra force can be expressed in terms of the baryonic mass, Newton's constant, and the Hubble acceleration scale $a_{0} = cH_{0}$, and provides evidence that this mechanism can account for the galactic and cluster-scale phenomena usually attributed to dark matter. Moreover, it is argued that both dark energy and the apparent dark matter effects originate from a common microscopic mechanism related to the emergent nature of spacetime and gravity. Empirically, the transition to this ``dark'' gravitational regime occurs when the gravitational acceleration drops below the critical scale, as first emphasized by Milgrom~\cite{Milgrom:1983ca, Milgrom:1983pn, Milgrom:1983zz}, which also corresponds to a characteristic surface mass density threshold observed in galactic dynamics.

In Hossenfelder’s covariant extension of Verlinde’s emergent gravity framework~\cite{Hossenfelder:2017eoh}, a generally covariant Lagrangian is constructed which allows for an improved interpretation of the underlying mechanism of emergent gravitational dynamics. In this model, de~Sitter space is effectively filled with a vector field that couples directly to baryonic matter and, through a ``dragging'' effect, mimics an additional gravitational force akin to dark matter. The covariant equation of motion is solved in a Schwarzschild background, yielding correction terms to the non-covariant expression first proposed by Verlinde. Furthermore, in this scheme, the same vector field can also reproduce the effect of dark energy, thereby offering a unified description of dark matter and dark energy phenomena within an emergent-gravity paradigm.

The thermodynamic interpretation of gravity thus offers a promising route to overcome the limitations of MOND. In particular, introducing temperature corrections to the entropic formulation of gravity allows one to retain MOND’s empirical successes at galactic scales while embedding them in a consistent relativistic and thermodynamic framework.

The original MOND proposal was purely non-relativistic and could not account for phenomena such as gravitational lensing or cosmological evolution. In this work, we seek to construct a \emph{relativistic extension of MOND} derived from entropic gravity with temperature-dependent corrections. Specifically, we aim to embed MOND within a modified Einstein framework in which the geometric sector of the field equations acquires thermal modifications. This approach not only provides a relativistic foundation for MOND but also establishes an explicit link between spacetime geometry and thermodynamic quantities such as entropy and temperature.

This work builds upon previous studies in both the entropic gravity and MOND literature. The foundational work of Verlinde~\cite{Verlinde:2010hp}, who proposed that gravity emerges from thermodynamic principles, provides a conceptual basis for our model. We extend his framework by introducing thermal corrections to the equipartition law, allowing for a more direct connection to MOND. Additionally, our work resonates with Milgrom's original MOND proposal~\cite{Milgrom:1983ca}, as well as later attempts to modify gravity in a relativistic context, such as those by Bekenstein~\cite{Bekenstein:2004ne} and Skordis and Zlosnik~\cite{Skordis:2020eui}, who explored relativistic versions of MOND. Our model also contributes to ongoing efforts to modify gravity in ways that do not require non-baryonic dark matter, echoing the ideas of Famaey and McGaugh \cite{Famaey:2011kh} and McGaugh~\cite{McGaugh:2011nv}, who have worked on explaining galaxy dynamics without invoking dark matter.

By embedding MOND in the context of entropic gravity, we provide a fresh perspective on the long-standing debate between modified gravity and dark matter. The model presented here offers a possible resolution to the mystery of dark matter by suggesting that the effects attributed to dark matter might instead be explained by modifications to gravity itself, particularly at large galactic scales.

To achieve this goal, we exploit and solve the modified Einstein equations for a static, spherically symmetric spacetime, obtaining the modified form of the metric in the regime of extremely weak gravitational fields. The resulting metric naturally dominates for the accelerations below a characteristic acceleration scale that provides a theoretical underpinning for the phenomenological constant $a_0$ in MOND. Finally, we confront our model with astrophysical data by comparing its predictions for galactic rotation curves with observations. In particular, we examine the galaxy NGC~3198 and compare the results of our relativistic MOND model with those from both the classical Newtonian dynamics (ND) and dark matter (DM) frameworks.

This paper is organized as follows. In Sec.~\ref{sec:temperature_correction}, we analyze the temperature corrections to the equipartition law of energy associated with information bits on the holographic screen. In Sec.~\ref{sec:Einstein}, we explore these corrections within the entropic force framework and investigate their consequences for the Einstein field equations. Sec.~\ref{sec:Bianchi} is devoted to examining the validity of the Bianchi identity in this context, along with its implications for energy conservation and thermodynamic interpretation. In Sec.~\ref{sec:metric}, we apply the modified gravitational equations to a static, spherically symmetric spacetime. In Sec.~\ref{sec:MOND}, we derive the MOND relation within our relativistic formulation. In Sec.~\ref{sec:obs}, we confront our theoretical predictions with observational data, showing that the framework can successfully explain galactic dynamics without the need for dark matter. Finally, Sec.~\ref{sec:con} presents a summary of our findings and outlines possible future directions.


\section{Temperature Correction to the Equipartition Law of Energy}
\label{sec:temperature_correction}

To analyze the temperature corrections to the Einstein equations, we apply the holographic principle to a given mass distribution. According to the holographic principle, all physical information confined within a spherical volume can be stored completely on its boundary surface in the form of discrete information bits. This view becomes especially meaningful when combined with the equipartition law of energy. When a particle passes through a spherical screen, its energy is distributed according to the energy partitioning law among all the information bits on the surface. The total number of degrees of freedom of the bits stored on the boundary of the system is given by~\cite{Verlinde:2010hp}
\begin{equation}
N=\frac{A}{\ell_{P}^{2}} \, ,
\label{N}
\end{equation}
where $A$ is the area of the two-dimensional boundary and $\ell_{P}=\sqrt{\hbar G/c^{3}}$ is the Planck length. Throughout the present work, the analysis is performed in natural units where $c=\hbar=k_{B}=1$. In Sec.~\ref{sec:obs}, where we compare the results of the theoretical model with experimental observations, it is useful to explicitly preserve the explicit form of these constants in the mathematical equations.

The entropy of a system is proportional to the number of degrees of freedom of the bits of information stored on its boundary,
\begin{equation}
S=\frac{N}{4}=\frac{A}{4\ell_{P}^{2}} \, .
\label{S}
\end{equation}
In his entropic force formulation to derive the Einstein gravitational field equations, Verlinde~\cite{Verlinde:2010hp} assumed that the energy of the bits of information stored on the holographic screen obeys the standard equipartition law of energy,
\begin{equation}
E = \frac{1}{2} N k_B T \, .
\label{E-standard}
\end{equation}
However, from the study of the heat capacity behavior of crystals in solid-state physics, we expect that the standard equipartition law of energy will be modified. In particular, the temperature corrections to the equipartition law have significant effects at low temperatures.

To explain the propagation of information in a space, it is assumed that holographic screens are foliated in space, adjacent to each other, so that each screen forms an equipotential surface~\cite{Verlinde:2010hp}. In fact, the bits of information on each of these plates can be considered as interacting fields that interact with each other, causing information to propagate from one screen to its neighbor. To describe the quantum energy of these interacting fields, we consider them as vibrating modes that can oscillate in different directions in space-time. The uncertainty principle of quantum mechanics, however, restricts the vibrations of these bits in the two dimensions tangent to the holographic screen. But these bits are still free to vibrate in the other two dimensions of space-time. We denote the number of degrees of freedom of the bits in these two dimensions by $N_1$ and $N_2$. As a result, the total number of degrees of freedom is equal to $N = N_1 + N_2$. Also, due to symmetry, it is logical to assume $N_1 = N_2$. The energy of these bits at high temperatures, which is associated with strong gravitational fields, is given by the standard equipartition law. It is expected that at low temperatures, associated with weak gravitational fields, the equipartition law will be modified. In the following, we want to obtain the modified equipartition law of energy in our setup. For this purpose, we apply the approach that Debye~\cite{Reif1965, Pathria:1996hda} used to derive a modified equation for the heat capacity of a crystal.

Assuming that the energy distribution of the oscillation modes follows the Bose-Einstein statistics, the occupancy number of the oscillation modes with frequency $\omega$ is given by the following equation~\cite{Pathria:1996hda}
\begin{equation}
n(\omega) = \frac{\hbar \omega}{e^{\hbar \omega/k_B T} - 1} \, .
\label{n-omega}
\end{equation}
Therefore, the total internal energy is obtained by integrating over all modes,
\begin{equation}
E = \int_0^{\omega_D} g(\omega) \, \frac{\hbar \omega}{e^{\hbar \omega / k_B T} - 1} \, d\omega \, ,
\label{E-omega}
\end{equation}
where $g(\omega)$ is the density of vibrational states.

Following Debye~\cite{Reif1965, Pathria:1996hda}, we assume that the vibrational spectrum is continuous up to the maximum cutoff frequency $\omega_D$, and for higher frequencies its contribution becomes negligible. Consequently, we have
\begin{equation}
\int_{0}^{\omega_{D}}g(\omega)\,d\omega=N \, ,
\end{equation}
For the Debye model, the density of states is given by
\begin{equation}
g(\omega)=\left\{ \begin{array}{cc}
\frac{N}{\omega_{D}^{2}} \, \omega \:, & 0\le\omega\le\omega_{D} \, ,
\\
0\,, & \omega>\omega_{D} \, .
\end{array}\right.
\label{g}
\end{equation}
Therefore, using this in Eq.~\eqref{E-omega}, the internal energy of the oscillators becomes
\begin{equation}
E=\frac{1}{2}Nk_{B}T\,D_{2}(T) \, .
\label{E-D2}
\end{equation}
Here, $D_{2}(x)$ indicates the two-dimensional Debye function which is defined as
\begin{align}
D_{2}(x) &= \frac{2}{x^{2}}\int_{0}^{x}\frac{t^{2}}{e^{t}-1}\,dt
\nonumber
\\
&= \frac{2}{x^{2}}\left[-2\text{Li}_{3}\left(e^{-x}\right)-2x\text{Li}_{2}\left(e^{-x}\right)+x^{2}\log\left(\sinh(x)-\cosh(x)+1\right)+2\zeta(3)\right] \, ,
\label{D2}
\end{align}
where $\text{Li}_{n}(z)$ is the polylogarithm function, and $\zeta(x)$ indicates the zeta function.

It is appropriate to define the dimensionless variable
\begin{equation}
x\equiv\frac{\hbar\omega_{D}}{k_{B}T}=\frac{T_{D}}{T} \, ,
\label{x}
\end{equation}
where $T_{D}\equiv\hbar\omega_{D}/k_{B}$ is the Debye temperature. According to the Unruh temperature formula~\eqref{T}, the Debye temperature is related to the Debye acceleration by $T_{D}=\hbar a_{D}/(2\pi ck_{B})$.

In the high temperature limit ($T\gg T_{D}$), the two-dimensional Debye function~\eqref{D2} can be approximated by
\begin{equation}
D_{2}(x)\approx1-\frac{x}{3}+\frac{x^{2}}{24}+\mathcal{O}(x^{3})\,,\qquad(x\ll1)
\label{D2-highT}
\end{equation}
Therefore, by ignoring the second and subsequent terms in the above equation compared to the first term, we see that at high temperatures, we arrive at the standard form of the energy equipartition law given in Eq.~\eqref{E-standard}. Also, we see in this equation that for $x \approx 3$, we have $D_{2}(x)$ is negligible compared to unity. This implies that for the temperatures $T\lesssim3T_{D}$ the energy equipartition law deviates considerably from its standard form and effectively enters the very low temperature regime. Equivalently, we expect that for the gravitational acceleration $a\lesssim3a_{D}$, the theory of gravity will be changed effectively, and we will enter the new regime of gravity theory. Hence, we define the transition acceleration as $a_{t}\equiv3a_{D}$, which indicates the effective threshold for changing the gravity theory.

At very low temperatures, the two-dimensional Debye function can be approximated as
\begin{equation}
D_{2}(x)\approx\frac{4\zeta(3)}{x^{2}}\,,\qquad(x\gg1) \, .
\label{D2-lowT}
\end{equation}
Therefore, in this regime, the equipartition law of energy~\eqref{E-D2} is completely different from the standard equipartition law. Therefore, we expect the resulting theory in this regime to be completely different from Einstein's general relativity. In the next section, we discuss the implications of the modified equipartition law of energy equipartition in the Verlinde entropic force approach.


\section{Modified Einstein Equations}
\label{sec:Einstein}

We consider the generalized equipartition law of the energy for the bits stored on the holographic screen as
\begin{equation}
E = \frac{1}{2} N k_B T f(T) \, .
\label{E}
\end{equation}
Here, $f(T)$ denotes the temperature correction function, which is a function of the temperature $T$ of the holographic screen. According to the discussion of the previous section, we saw that this function can be considered equal to the two-dimensional Debye function. Here, we treat this function as a general function, and only in the final parts of Sec.~\ref{sec:metric}, we set it approximately equal to the two-dimensional Debye function.

Using the mass-energy equivalence $E=M$, we define the total thermodynamic mass $M_{\rm th}$ enclosed by the screen $\mathcal{S}$ as an integral over its area,
\begin{equation}
M_{\rm th} = \int dM = \int_\mathcal{S} \left( \frac{1}{2G\hbar} T f(T) \right) dA \, .
\label{Mth}
\end{equation}
To express this covariantly, we identify the temperature $T$ with the Unruh temperature~\cite{Unruh:1976db} for a static observer following a timelike Killing vector field $\xi^a$. This temperature is given by
\begin{equation}
T = \frac{\hbar a}{2\pi} \, ,
\label{T}
\end{equation}
where the acceleration satisfies $a = e^\phi N^b \nabla_b \phi$. Here, $\phi = \frac{1}{2}\ln(-\xi^a\xi_a)$ is the redshift potential and $N^b$ is the unit normal to the equipotential screen $\mathcal{S}$.

Substituting this covariant temperature $T = \frac{\hbar}{2\pi} e^\phi N^b \nabla_b \phi$ gives our final expression for the thermodynamic mass~\cite{Sheykhi:2012vf, Rezazadeh:2025htb},
\begin{equation}
M_{\rm th} = \int_\mathcal{S} \left( \frac{1}{2G\hbar} \right) \left( \frac{\hbar}{2\pi} e^\phi N^b \nabla_b \phi \right) f(T) dA \, .
\label{Mth2}
\end{equation}
This equation is the thermodynamic definition of the mass enclosed by the screen, modified by the general statistical function $f(T)$.

Next, we seek a purely geometric definition of mass, $M_{\rm geo}$, to equate with $M_{\rm th}$. In the standard general relativity, the Komar mass $M_K$ is given by~\cite{Wald:1984rg}
\begin{equation}
M_K = -\frac{1}{8\pi G} \int_\mathcal{S} \nabla^a \xi^b d\mathcal{S}_{ab} \, .
\label{MK}
\end{equation}
This integral is equivalent to
\begin{equation}
M_K = \frac{1}{4\pi G} \int_\mathcal{S} (e^\phi N^b \nabla_b \phi) dA\, .
\label{MK2}
\end{equation}
Comparing $M_K$ from Eq. \eqref{MK2} with our $M_{\rm th}$ in Eqs.~\eqref{Mth2}, we see that
\begin{equation}
M_{\rm th} = \int_S f(T) dM_K \, .
\label{Mth-MK}
\end{equation}
This result motivates our central postulate: The correct geometric mass $M_{\rm geo}$ in this generalized entropic gravity model is a modified Komar integral, where the scalar correction function $f(T)$ is inserted into the integrand,
\begin{equation}
M_{\rm geo} \equiv -\frac{1}{8\pi G} \int_\mathcal{S} f(T) \nabla^a \xi^b d\mathcal{S}_{ab} \, .
\label{Mgeo}
\end{equation}

We now use the generalized Stokes' theorem
\begin{equation}
\oint_S \omega = \int_\Sigma d\omega \, ,
\label{Stokes}
\end{equation}
to transform this 2-surface integral (over $\mathcal{S}$) into a 3-volume integral (over the enclosed hypersurface $\Sigma$). For an antisymmetric tensor field $B^{ab}$, this theorem indicates~\cite{poisson2004relativist}
\begin{equation}
\oint_S B^{ab} d\mathcal{S}_{ab} = 2 \int_\Sigma \nabla_b B^{ab} d\Sigma_a \, .
\label{Stokes2}
\end{equation}
Applying this to $M_{\rm geo}$ in Eq.~\eqref{Mgeo} with $B^{ab} = f(T) \nabla^a \xi^b$, we find
\begin{equation}
M_{\rm geo} = -\frac{1}{4\pi G}  \int_\Sigma \nabla_b \left( f(T) \nabla^a \xi^b \right) d\Sigma_a \, .
\label{Mgeo2}
\end{equation}
Using the product rule for the covariant derivative, we arrive at
\begin{equation}
M_{\rm geo} = -\frac{1}{4\pi G} \int_\Sigma \left[ (\nabla_b D(x)) (\nabla^a \xi^b) + f(T) (\nabla_b \nabla^a \xi^b) \right] d\Sigma_a \, .
\label{Mgeo3}
\end{equation}
We simplify the second term using two identities for the Killing vector $\xi^a$~\cite{Wald:1984rg},
\begin{align}
\nabla_b \nabla^a \xi^b &= -\nabla_b \nabla^b \xi^a \, ,
\\
\nabla_b \nabla^b \xi^a &= -R^a_c \xi^c \, .
\end{align}
These imply
\begin{equation}
\nabla_b \nabla^a \xi^b = R^a_c \xi^c \, .
\label{xi-R}
\end{equation}
Substituting this into our integral~\eqref{Mgeo3} for the geometrical mass, we obtain
\begin{equation}
M_{\rm geo} = -\frac{1}{4\pi G} \int_\Sigma \left[ \nabla^a \xi^b \nabla_b f(T) + f(T) R^a_c \xi^c \right] d\Sigma_a \, .
\label{Mgeo4}
\end{equation}
For a spacelike hypersurface $\Sigma$, the directed surface element is $d\Sigma_a = -n_a d\Sigma$, where $n^a$ is the future-pointing unit normal. Consequently, the above equation turns into
\begin{equation}
M_{\rm geo} = \frac{1}{4\pi G} \int_\Sigma \left[ f(T) R_{ac} \xi^c + \nabla_a \xi^c \nabla_c f(T) \right] n^a d\Sigma \, .
\label{Mgeo5}
\end{equation}
This expression represents the total mass of the spacetime purely in terms of its generalized, statistically-modified geometry.

The final step is to equate this geometric definition of mass $M_{\rm geo}$ with the physical definition of mass derived from the stress-energy tensor $\mathcal{T}_{ab}$. This matter-energy mass $M_{\rm matter}$ is given by~\cite{Wald:1984rg}
\begin{equation}
M_{\rm matter} = 2 \int_\Sigma \left( \mathcal{T}_{ab} - \frac{1}{2} \mathcal{T} g_{ab} \right) n^a \xi^b d\Sigma \, .
\label{Mmatter}
\end{equation}
The fundamental principle of gravity, $M_{\rm geo} = M_{\rm matter}$, implies
\begin{equation}
\frac{1}{4\pi G} \int_\Sigma \left[ f(T) R_{ac} + \nabla_a \xi^c \nabla_c f(T) g_{bc} \right] n^a \xi^c d\Sigma = 2 \int_\Sigma \left( \mathcal{T}_{ac} - \frac{1}{2} \mathcal{T} g_{ac} \right) n^a \xi^c d\Sigma \, ,
\label{Mgeo-Mmatter}
\end{equation}
where have relabeled indices $b \to c$ for direct comparison. Since this equivalence must hold for any arbitrary integration volume $\Sigma$, any time-translation Killing field $\xi^c$, and any normal field $n^a$, the tensor-valued integrands must be equal. This yields the generalized Einstein field equations~\cite{Sheykhi:2012vf, Rezazadeh:2025htb}
\begin{equation}
f(T) R_{ab} - e^{-2\phi}\xi_{b}\nabla_{a}\xi^{c}\nabla_{c} f(T) = 8\pi G \left( \mathcal{T}_{ab} - \frac{1}{2}\mathcal{T} g_{ab} \right) \, .
\label{Einstein}
\end{equation}
This is the most general form of the modified field equations, emerging from the premise that the holographic principle is subject to a non-classical statistical correction $f(T)$. In the limit of high temperatures, which correspond to strong gravitational fields, we have $f(T) = 1$, and the above equation reduces to the standard Einstein equations. But at low temperatures, which correspond to the regime of very weak gravitational fields, the effects of temperature corrections may cause the theory of gravity to be quite different from Einstein's general relativity.


\section{Bianchi Identities, Non-Conservation Law, and Thermodynamic Interpretation}
\label{sec:Bianchi}

In this section, we investigate the consistency and physical implications of the modified gravitational field equations derived in Eq.~\eqref{Einstein}, which deviate from the standard Einstein equations through the presence of a temperature-dependent function $f(T)$. A central question is whether these generalized equations remain compatible with fundamental geometric identities, in particular the contracted Bianchi identity, and how they affect the conservation of the energy-momentum tensor. By taking the covariant divergence of the field equations, we show that the usual conservation law is modified and replaced by a generalized balance equation involving an effective exchange current. This indicates that energy and momentum are not conserved independently within the matter sector but are exchanged with additional gravitational or thermodynamic degrees of freedom. We further interpret this structure within a thermodynamic framework by introducing a generalized entropy functional and invoking the Clausius relation, thereby establishing a connection between the modified dynamics and non-equilibrium gravitational thermodynamics, with general relativity emerging as the equilibrium limit.

Since the field equations~\eqref{Einstein} differ from the standard Einstein equations, it is important to analyze their compatibility with the contracted Bianchi identities and the corresponding conservation laws. Taking the covariant divergence of Eq.~\eqref{Einstein}, we obtain
\begin{equation}
\nabla^{a}\left[f(T)R_{ab}\right]-\nabla^{a}\left[e^{-2\phi}\,\xi_{b}\nabla_{a}\xi^{c}\nabla_{c}f(T)\right]=8\pi G\nabla^{a}\left(\mathcal{T}_{ab}-\frac{1}{2} \mathcal{T} g_{ab}\right) \, .
\label{cov-Einstein}
\end{equation}
Using the geometric identity
\begin{equation}
\nabla^a R_{ab}=\frac12\nabla_b R \, ,
\label{cov-R}
\end{equation}
the first term in Eq.~\eqref{cov-Einstein} becomes
\begin{equation}
\nabla^a(fR_{ab}) = (\nabla^a f)R_{ab} + \frac12 f\nabla_b R \, .
\tag{cov-fR}
\end{equation}

Consequently, the matter energy-momentum tensor satisfies the generalized conservation equation,
\begin{equation}
\nabla^a \mathcal{T}_{ab}=J_b \, ,
\label{cov-T}
\end{equation}
where the exchange current $J_b$ is given by
\begin{align}
J_{b}= & \frac{1}{8\pi G}\Big[(\nabla^{a}f)R_{ab}+\frac{1}{2}f\nabla_{b}R-e^{-2\phi}\Big(-2(\nabla^{a}\phi)\,\xi_{b}\nabla_{a}\xi^{c}\nabla_{c}f
\nonumber
\\
& +(\nabla^{a}\xi_{b})\,\nabla_{a}\xi^{c}\nabla_{c}f+\xi_{b}(\Box\xi^{c})\,\nabla_{c}f+\xi_{b}\nabla_{a}\xi^{c}\nabla_{c}\nabla^{a}f\Big)\Big]+\frac{1}{2}\nabla_{b}\mathcal{T} \, .
\label{Jb}
\end{align}

Equation~\eqref{cov-T} indicates that the matter sector is not conserved independently. Instead, the modified dynamics describe an exchange of energy and momentum between matter and an effective thermal/gravitational sector encoded in the temperature-dependent function $f(T)$, the scalar field $\phi$, and the vector field $\xi^a$.

This structure is analogous to non-equilibrium thermodynamic descriptions of gravity and interacting gravitational systems. The contracted Bianchi identity remains satisfied at the level of the effective theory,
\begin{equation}
\nabla^a G_{ab}=0 \, ,
\label{cov-Gab}
\end{equation}
while the non-vanishing current $J_b$ represents an effective transfer current between spacetime and matter degrees of freedom.

To make this interpretation explicit, we introduce the generalized entropy functional
\begin{equation}
S[f(T)]=\frac{1}{4G}\int_{\Sigma}f(T)e^{-2\phi}\,dA \, ,
\label{S-f}
\end{equation}
where $\Sigma$ denotes a local horizon cross-section and $dA$ is the induced area element. In the limit $f(T)=1$ and $\phi=0$, Eq.~\eqref{S-f} reduces to the standard Bekenstein--Hawking entropy,
\begin{equation}
S_{\rm BH}=\frac{A}{4G} \, .
\label{S-BH}
\end{equation}

The variation of the entropy functional~\eqref{S-f} contains additional contributions proportional to gradients of $f(T)$,
\begin{equation}
\delta S=\frac{1}{4G}\int_{\Sigma}\left[f(T)\,\delta(dA)+dA\,\delta f(T)\right] \, .
\label{deltaS}
\end{equation}

Following Jacobson's thermodynamic derivation of gravitational dynamics~\cite{Jacobson:1995ab}, we impose the local Clausius relation,
\begin{equation}
\delta Q=T\,\delta S \, .
\label{Clausius}
\end{equation}
The resulting entropy variation generates both the curvature term $f(T)R_{ab}$ and additional gradient contributions involving $f(T)$, $\phi$, and $\xi^a$. The modified field equations may therefore be interpreted as effective gravitational equations associated with a spacetime-dependent entropy density.

Projecting Eq.~\eqref{Jb} along a timelike observer $u^b$, one obtains
\begin{equation}
u^b J_b=T\nabla_a s^a \, ,
\label{ub-Jb}
\end{equation}
where $s^a$ denotes the entropy current. The exchange current $J_b$ therefore measures local entropy production and energy transfer between matter and spacetime degrees of freedom.

In equilibrium configurations characterized by constant $f(T)$, stationary geometry, and vanishing entropy flux, one recovers
\begin{equation}
J_b = 0 \, ,
\label{Jb-equilibrium}
\end{equation}
and the standard covariant conservation law holds,
\begin{equation}
\nabla^a \mathcal{T}_{ab}=0 \, .
\label{cov-T-equilibrium}
\end{equation}
This implies that standard general relativity arises as the equilibrium limit of the present thermodynamic framework.


\section{Solution of Modified Einstein Equations}
\label{sec:metric}

In this section, we are primarily interested in the solution of the modified Einstein equations \eqref{Einstein} for a static, spherically symmetric space-time with the line element
\begin{equation}
ds^2 = -A(r) dt^2 + B(r) dr^2 + r^2 (d\theta^2 + \sin^2\theta \, d\phi^2) \, .
\label{metric}
\end{equation}
In this equation, $A(r)$ and $B(r)$ represent the metric coefficients, which are functions of the radial coordinate $r$. Our goal in the following is to determine the analytical form of these coefficients by solving the modified Einstein equations.

The Christoffel symbols of the second kind are given by the following equation
\begin{equation}
\Gamma_{\mu\nu}^{\lambda}=\frac{1}{2}g^{\lambda\sigma}\left(\partial_{\mu}g_{\nu\sigma}+\partial_{\nu}g_{\mu\sigma}-\partial_{\sigma}g_{\mu\nu}\right) \, .
\label{Gamma}
\end{equation}
Using this, the non-zero Christoffel symbols for the metric~\eqref{metric} are obtained as
\begin{align}
\Gamma_{tr}^{t} &= \Gamma_{rt}^{t}=\frac{A'}{2A} \, ,
\label{Gammat}
\\
\Gamma_{tt}^{r} &= \frac{A'}{2B},\quad\Gamma_{rr}^{r}=\frac{B'}{2B},\quad\Gamma_{\theta\theta}^{r}=-\frac{r}{B},\quad\Gamma_{\phi\phi}^{r}=-\frac{r\sin^{2}\theta}{B} \, ,
\label{Gammar}
\\
\Gamma_{r\theta}^{\theta} &= \Gamma_{\theta r}^{\theta}=\frac{1}{r},\quad\Gamma_{\phi\phi}^{\theta}=-\sin\theta\cos\theta \, ,
\label{Gammatheta}
\\
\Gamma_{r\phi}^{\phi} &= \Gamma_{\phi r}^{\phi}=\frac{1}{r},\quad\Gamma_{\theta\phi}^{\phi}=\Gamma_{\phi\theta}^{\phi}=\cot\theta \, .
\label{Gammaphi}
\end{align}
Now, we can calculate the Riemann tensor, which is given by the following equation
\begin{equation}
R_{\sigma\mu\nu}^{\rho}=\partial_{\mu}\text{\ensuremath{\Gamma}}_{\nu\sigma}^{\rho}-\partial_{\nu}\text{\ensuremath{\Gamma}}_{\mu\sigma}^{\rho}+\text{\ensuremath{\Gamma}}_{\mu\lambda}^{\rho}\text{\ensuremath{\Gamma}}_{\nu\sigma}^{\lambda}-\text{\ensuremath{\Gamma}}_{\nu\lambda}^{\rho}\text{\ensuremath{\Gamma}}_{\mu\sigma}^{\lambda} \, .
\label{Reimann}
\end{equation}
The Ricci tensor is obtained by contracting the Riemann tensor through $R_{\mu\nu}\equiv R_{\mu\lambda\nu}^{\lambda}$. The non-zero components of the Ricci tensor for the metric \eqref{metric} are as follows
\begin{align}
R_{tt} &= \frac{A''}{2B}-\frac{A'B'}{4B^{2}}-\frac{A'^{2}}{4AB}+\frac{A'}{rB} \, ,
\label{Rtt}
\\
R_{rr} &= -\frac{A''}{2A}+\frac{A'B'}{4AB}+\frac{A'^{2}}{4A^{2}}+\frac{B'}{rB} \, ,
\label{Rrr}
\\
R_{\theta\theta} &= 1-\frac{rA'}{2AB}+\frac{rB'}{2B^{2}}-\frac{1}{B} \, ,
\label{Rthetatheta}
\\
R_{\phi\phi} &= \sin^{2}\theta\,R_{\theta\theta}=\sin^{2}\theta\left(1-\frac{rA'}{2AB}+\frac{rB'}{2B^{2}}-\frac{1}{B}\right) \, .
\label{Rphiphi}
\end{align}
The Ricci tensor can be further contracted via $R\equiv R_{\mu}^{\mu}$ to give the Ricci scalar as
\begin{equation}
R=-\frac{A''}{AB}+\frac{A'B'}{2AB^{2}}+\frac{A'^{2}}{2A^{2}B}+\frac{2B'}{rB^{2}}-\frac{2A'}{rAB}+\frac{2}{r^{2}}-\frac{2}{r^{2}B} \, .
\label{R}
\end{equation}

We are looking for the solution to the gravitational field equation in a vacuum. Therefore, the stress-energy tensor vanishes
\begin{equation}
\mathcal{T}^{\mu}_{\nu} = 0 \, .
\label{Tmunu0}
\end{equation}
As a result, the modified Einstein equations~\eqref{Einstein} take the following form
\begin{equation}
f(T)R_{\mu\nu}-e^{-2\phi}\text{\ensuremath{\xi}}_{\nu}\nabla_{\mu}\xi^{\lambda}\partial_{\lambda}f(T)=0 \, .
\label{Einstein-Tmunu0}
\end{equation}

With our complete characterization of the spacetime geometry and explicit Christoffel symbols at hand, we can now proceed to determine the Killing vector in our setting. For this purpose, we should find an appropriate solution to the Killing equation,
\begin{equation}
\nabla_\mu \xi_\nu + \nabla_\nu \xi_\mu = 0 \, ,
\label{Killing}
\end{equation}
which provides all the required symmetries. Using the covariant derivative
\begin{equation}
\nabla_\mu \xi_\nu = \partial_\mu \xi_\nu - \Gamma^\lambda_{\mu\nu} \xi_\lambda \, ,
\label{covDxi}
\end{equation}
we arrive at
\begin{equation}
\partial_\mu \xi_\nu + \partial_\nu \xi_\mu - 2 \Gamma^\lambda_{\mu\nu} \xi_\lambda = 0 \, ,
\label{pDxi}
\end{equation}
where the covariant Killing vector is given by
\begin{equation}
\xi_{\mu}=g_{\mu\nu}\xi^{\nu}=\left(-A\xi^{t},B\xi^{r},r^{2}\xi^{\theta},r^{2}\sin^{2}\theta\,\xi^{\phi}\right) \, .
\label{covxi}
\end{equation}
The metric's symmetries yield four sets for the Killing vectors:\\
\noindent 1. Time translation: Owing to the $t$-independence of the metric \eqref{metric}, the vector $\xi^\mu = (1, 0, 0, 0)$ satisfies all Killing equations~\eqref{pDxi}, confirming that it is a Killing vector.

\noindent 2. Rotation around the $z$-axis: Because the metric is independent of $\phi$, the vector $\xi^\mu = (0, 0, 0, 1)$ satisfies all the equations~\eqref{pDxi}, and thus represents another Killing vector.

\noindent 3. Rotation around the $x$-axis: Due to the spherical symmetry of the metric, the vector $\xi^\mu = (0, 0, \sin\phi, \cot\theta \cos\phi)$ also satisfies the Killing equations~\eqref{pDxi}, corresponding to a rotation around the $x$-axis.

\noindent 4. Rotation around the $y$-axis: Similarly, the vector $\xi^\mu = (0, 0, -\cos\phi, \cot\theta \sin\phi)$ fulfills Eq.~\eqref{pDxi}, representing a rotation around the $y$-axis.

In this section, we were able to extract valuable insights regarding the symmetries of the problem. These symmetries not only help us understand the underlying structure of the space-time but also simplify the subsequent analysis. However, the main focus of our analytical computations will be devoted to the case of \textit{time translation symmetry}, which plays a fundamental role in the conservation of energy.

We now have all the ingredients required to evaluate the field equations explicitly. Imposing time-translation symmetry, we substitute the Killing vector $\xi^\mu = (1,0,0,0)$ into Eq.~\eqref{Einstein-Tmunu0}. For the static, spherically symmetric metric~\eqref{metric}, the gravitational acceleration depends only on the radial coordinate $r$. Consequently, the associated Unruh temperature $T$ is also a function of $r$ alone. This temperature characterizes the effective temperature of the holographic screen associated with a static observer and should not be interpreted as a thermodynamic temperature of matter or radiation. As a result, the correction function $f(T)$ depends only on the radial coordinate $r$. With these considerations, the components of Einstein's equations~\eqref{Einstein-Tmunu0} are obtained as follows
\begin{align}
& f(r)\left(\frac{A''}{2B}-\frac{A'B'}{4B^{2}}-\frac{A'^{2}}{4AB}+\frac{A'}{rB}\right)+\frac{A'f'(r)}{2B}=0 \, ,
\label{Einstein00}
\\
& f(r)\left(-\frac{A''}{2A}+\frac{A'B'}{4AB}+\frac{A'^{2}}{4A^{2}}+\frac{B'}{rB}\right)=0 \, ,
\label{Einstein11}
\\
& f(r)\left(1-\frac{rA'}{2AB}+\frac{rB'}{2B^{2}}-\frac{1}{B}\right)=0 \, .
\label{Einstein22}
\end{align}
Combining Eqs. \eqref{Einstein00} and \eqref{Einstein11} to eliminate $A''$, we reach the following equation
\begin{equation}
BA'\left(rf'+2f\right)+2AfB'=0 \, .
\label{Einstein1}
\end{equation}
Also, it is obvious that Eq.~\eqref{Einstein22} can be simplified to
\begin{equation}
1-\frac{rA'}{2AB}+\frac{rB'}{2B^{2}}-\frac{1}{B}=0 \, .
\label{Einstein2}
\end{equation}
Eqs.~\eqref{Einstein1} and \eqref{Einstein2} are the two main equations in our work to determine the two coefficients $A(r)$ and $B(r)$ in the metric~\eqref{metric}.

If the form of the function $f$ is explicitly known in terms of $r$, then Eqs.~\eqref{Einstein1} and \eqref{Einstein2} determine the metric coefficients as follows
\begin{align}
A(r) &= C_{1}\exp\left[-4e^{C_{2}}\int\frac{f(r)}{r\left(rf'(r)+4f(r)\right)\left[e^{C_{2}}+\exp\left(\int\frac{2rf'(r)+4f(r)}{r^{2}f'(r)+4rf(r)}\,dr\right)\right]}\,dr\right] \, ,
\label{A-fr}
\\
B(r) &= \frac{1}{\exp\left[C_{2}-\int\frac{2rf'(r)+4f(r)}{r^{2}f'(r)+4rf(r)}\,dr\right]+1} \, ,
\label{B-fr}
\end{align}
where $C_1$ and $C_2$ are constant parameters. Since in our work, the explicit dependence of the function $f$ on $r$ is not known, the above two equations are not applicable in our approach, and we have to use the differential equations~\eqref{Einstein1} and \eqref{Einstein2} directly.

In our model, the function $f$ is a function of gravitational acceleration $a$. To determine gravitational acceleration, we should use the geodesic equations, which are given generally by
\begin{equation}
\frac{\partial^{2}x^{\mu}}{\partial\tau^{2}}+\text{\ensuremath{\Gamma}}_{\nu\lambda}^{\mu}\frac{\partial x^{\text{\ensuremath{\lambda}}}}{\partial\tau}\frac{\partial x^{\text{\ensuremath{\nu}}}}{\partial\tau}=0 \, .
\label{geodesic}
\end{equation}
For the metric~\eqref{metric}, the above equation leads to the following equations
\begin{align}
& \frac{\partial^{2}t}{\partial\tau^{2}}+\frac{A'}{A}\frac{\partial t}{\partial\tau}\frac{\partial r}{\partial\tau}=0 \, ,
\label{geodesic0}
\\
& \frac{\partial^{2}r}{\partial\tau^{2}}+\frac{A'}{2B}\left(\frac{\partial t}{\partial\tau}\right)^{2}+\frac{B'}{2B}\left(\frac{\partial r}{\partial\tau}\right)^{2}-\frac{r}{B}\left(\frac{\partial\theta}{\partial\tau}\right)^{2}-\frac{r\sin^{2}\theta}{B}\left(\frac{\partial\phi}{\partial\tau}\right)^{2}=0 \, ,
\label{geodesic1}
\\
& \frac{\partial^{2}\theta}{\partial\tau^{2}}-\sin\theta\cos\theta\left(\frac{\partial\phi}{\partial\tau}\right)^{2}+\frac{2}{r}\left(\frac{\partial r}{\partial\tau}\right)\left(\frac{\partial\theta}{\partial\tau}\right)=0 \, ,
\label{geodesic2}
\\
& \frac{\partial^{2}\phi}{\partial\tau^{2}}+2\cot\theta\left(\frac{\partial\theta}{\partial\tau}\right)\left(\frac{\partial\phi}{\partial\tau}\right)+\frac{2}{r}\left(\frac{\partial r}{\partial\tau}\right)\left(\frac{\partial\phi}{\partial\tau}\right)=0 \, .
\label{geodesic3}
\end{align}

In our analysis, we deal with very weak gravitational fields and low velocities that are significantly lower than the speed of light. Taking these assumptions into account, we can ignore the second term on the left side of Eq.~\eqref{geodesic0} compared to the first term~\cite{weinberg2013gravitation}. Similarly, in \eqref{geodesic1}, only the first two terms remain on the left side, and the other terms can be neglected in front of them~\cite{weinberg2013gravitation}. Therefore, Eqs.~\eqref{geodesic0} and \eqref{geodesic1} can be rewritten approximately as
\begin{align}
& \frac{\partial^{2}t}{\partial\tau^{2}}\approx0 \, .
\label{geodesic0approx}
\\
& \frac{\partial^{2}r}{\partial\tau^{2}}+\frac{A'}{2B}\left(\frac{\partial t}{\partial\tau}\right)^{2}\approx0 \, .
\label{geodesic1approx}
\end{align}
Using these equations, the gravitational acceleration is obtained in our model as
\begin{equation}
a=\frac{\partial^{2}r}{\partial t^{2}}\approx-\frac{A'}{2B} \, .
\label{a}
\end{equation}
This is the general form of the gravitational acceleration for the static, spherically metric~\eqref{metric} in the regime of weak gravitational fields and low velocities.

In the following, we consider the temperature correction function equal to the two-dimensional Debye function,
\begin{equation}
f(T)=D_{2}(T) \, .
\label{f-D2}
\end{equation}
If we consider the two-dimensional Debye function in its full form given in Eq.~\eqref{D2}, then the two Eqs.~\eqref{Einstein1} and \eqref{Einstein2} lead to very complicated equations that cannot be solved analytically. In this case, it is necessary to use numerical methods to solve the equations, which is a complex project that is beyond the scope of the present analysis work and is left to future investigations. However, to resolve this difficulty, we utilize in our work the approximation of the two-dimensional Debye function in the low-temperature regime, as given in Eq.~\eqref{D2-lowT}. Of course, it is necessary to note that Eq.~\eqref{D2-lowT} is fully valid only when a particle in the vicinity of the holographic screen feels only the gravitational effects of the mass enclosed within the screen, and it is also necessary that the particle's distance from that mass distribution is far enough away for the low-temperature limit to hold thoroughly. But our goal is to study the dynamics of stars at the edge of a galaxy, which, first, may not be far enough from the galactic center where a significant portion of the galaxy's mass is concentrated. Also, this star may feel the gravitational effects of nearby galaxies in addition to the effects of the host galaxy on the curvature of space-time. To include these effects into account, we consider a small correction to Eq.~\eqref{D2-lowT} and write it as
\begin{equation}
D_{2}(x)\approx\frac{4\zeta(3)}{x^{2+\delta}} \, ,
\label{D2approx-delta}
\end{equation}
where $\delta$ is a constant parameter whose value is much smaller than unity. We will determine this parameter in the following.

Now, we can show that the following solutions satisfy the modified Einstein equations~\eqref{Einstein1} and \eqref{Einstein2},
\begin{align}
A(r) &= 1+C_{A}r^{\epsilon} \, ,
\label{A}
\\
B(r) &= \frac{(2+\epsilon)\left(1+C_{A}r^{\epsilon}\right)}{2\left(1+C_{A}r^{\epsilon}\right)+\epsilon} \, ,
\label{B}
\end{align}
where $C_A$ is the integration constant and $\epsilon$ is a constant parameter whose value is much smaller than unity. We will see later that this parameter is of the same order as the $\delta$ parameter in Eq.~\eqref{D2approx-delta}. Also, the temperature correction function is determined in our scenario as
\begin{equation}
f(r)=\tilde{C}\left(2+2C_{A}r^{\epsilon}+\epsilon\right)^{\frac{2}{2+\epsilon}}r^{-\frac{4(1+\epsilon)}{2+\epsilon}} \, ,
\label{f}
\end{equation}
where $\tilde{C}$ is a constant parameter. Eqs.~\eqref{A} and \eqref{B} are the solutions of the modified Einstein equations in our relativistic MOND scenario that determine the curvature of space-time in the very weak gravitational regime. In deriving these equations, we have imposed the condition of asymptotically flatness for both the metric coefficients. Therefore, these coefficients satisfy the asymptotic behaviors $A(r) \approx 1$ and $B(r) \approx 1$ in the limit of very weak gravitational fields and large radii. They are not intended to describe the spacetime near $r=0$ or in strong-field regimes. Thus, the smallness of the parameter $\epsilon$ is consistent within the intended domain of validity.

Considering these conditions, Eq.~\eqref{f} can be approximated as
\begin{equation}
f(r)\approx\tilde{C}\left(2+\epsilon\right)^{\frac{2}{2+\epsilon}}r^{-\frac{4(1+\epsilon)}{2+\epsilon}} \, ,
\label{fapprox}
\end{equation}
where $\tilde{C}$ is a constant parameter. On the other hand, we assume the asymptotic form~\eqref{D2approx-delta}, we obtain this function as
\begin{equation}
f(r)\approx2^{-(2+\delta)}Ca_D^{-\delta-2}C_{A}^{\delta+2}\epsilon^{\delta+2}r^{(\delta+2)(\epsilon-1)} \, .
\label{fapprox-delta}
\end{equation}
By equating Eqs.~\eqref{fapprox} and \eqref{fapprox-delta}, the parameters $\delta$ and $\tilde{C}$ are determined as
\begin{align}
\delta & = \frac{2\epsilon(3+\epsilon)}{2-\epsilon-\epsilon^{2}}\approx3\epsilon \, ,
\label{delta}
\\
\tilde{C} & =2^{\frac{2\epsilon(\epsilon+3)}{-2+\epsilon+\epsilon^{2}}}(2+\epsilon)^{-\frac{2}{2+\epsilon}}\zeta(3)\left(\frac{C_{A}\epsilon}{a_{D}}\right)^{\frac{4(1+\epsilon)}{2-\epsilon-\epsilon^{2}}}\approx\frac{\zeta(3)}{2}\left(\frac{C_{A}\epsilon}{a_{D}}\right)^{2} \, .
\label{Ctilde}
\end{align}
In Eq.~\eqref{delta}, it is clear that the parameter $\delta$ is of the order of the parameter $\epsilon$, which is much smaller than unity.

By using Eqs.~\eqref{A} and \eqref{B} in Eq.~\eqref{a}, the magnitude of the gravitational acceleration is obtained as
\begin{equation}
|a|\approx\frac{1}{2}|C_{A} \epsilon|r^{-1+\epsilon} \, .
\label{absa}
\end{equation}
The magnitude of the gravitational acceleration is related to the rotational velocity by $|a|=v^{2}/r$. Consequently, we find that the rotational velocity of a test particle in the vicinity of the holographic screen in our RMOND model is given by
\begin{equation}
v=\sqrt{r|a|}\approx\sqrt{\frac{|C_{A} \epsilon|}{2}}r^{\epsilon/2} \, .
\label{vRMOND-a}
\end{equation}
In Sec.~\ref{sec:obs}, we use the above equation to estimate the rotational velocities of stars located at far distances from the center of the galaxy and compare the results of our model with the observational results.


\section{Derivation of the MOND Interpolation Function}
\label{sec:MOND}

An important test that requires further calculation and analysis is to examine to what extent the acceleration introduced in our framework \eqref{absa} can reproduce the standard MOND theory. This naturally leads to the question of why MOND theory is of particular importance.

The phenomenological success of the MOND paradigm in explaining galactic kinematics, most notably the flat rotation curves and the baryonic Tully-Fisher relation, without invoking cold dark matter, remains a crucial benchmark for any alternative theory of gravity. A fundamental challenge for theoretical frameworks derived from first principles, such as our model, is to naturally recover this empirical behavior at large astrophysical scales. Therefore, it is of utmost importance to investigate whether our derived weak-field metric perfectly aligns with the standard MOND formalism in the ultra-low acceleration regime.

Although standard MOND theory successfully explains galactic rotation curves, its fundamental theoretical origin is often introduced phenomenologically. Thus, it is well-motivated to establish whether fundamental gravitational modifications can naturally yield the MOND interpolation function. In this section, we show that the MOND interpolation function $\mu$ can be explicitly extracted from our modified metric formulation.

The MOND framework may be interpreted as an empirical modification of Newtonian gravity, wherein the physical acceleration, $|a|$, is replaced by an effective acceleration governed by
\begin{equation}
|a| \mu \left(\frac{|a|}{a_{0}}\right) = a_{N} \equiv \frac{GM}{r^{2}} \, .
\label{MOND}
\end{equation}
Here, $a_{N}$ is called the Newtonian acceleration. Furthermore, the function $\mu\left(\frac{|a|}{a_{0}}\right)$ is referred to as the MOND interpolation function and satisfies the following conditions
\begin{equation}
\mu\left(\frac{|a|}{a_{0}}\right)=\left\{\begin{array}{cc}
1\,, & a\gg a_{0}\,,\\
\frac{|a|}{a_{0}}\,, & a\ll a_{0}\,.
\end{array}\right.
\label{mu}
\end{equation}
The parameter $a_0$ is known as the Milgrom acceleration and is of order $10^{-10}\:m/s^{2}$. We want to show that in our relativistic framework, we can recover this formula and provide an approximate relation for the function $\mu\left(\frac{|a|}{a_{0}}\right)$. To this end, we note that in our model for accelerations greater than the transition acceleration ($a \gtrsim a_{t}$), the Einstein gravity regime holds, but for the accelerations less than it ($a \lesssim a_{t}$), our model deviates from the standard general relativity.

Here, it is useful to introduce the effective interpolation function
\begin{equation}
\mu_{\mathrm{eff}}\left(|a|\right)\equiv\frac{a_{N}}{|a|} \, .
\label{mueff}
\end{equation}
Our goal is to calculate this function in our model and show that it approximately satisfies the conditions~\eqref{mu} for the MOND function.

Combining Eqs.~\eqref{Einstein1}, \eqref{Einstein1}, and \eqref{a}, the gravitational acceleration relation in our scenario is given by
\begin{equation}
a=-\frac{2Af(B-1)}{rB\left(rf'+4f\right)} \, .
\label{a-f}
\end{equation}

In our model, the gravitational theory changes around the transition acceleration $a_t$, we show that it is of the order of the Milgrom acceleration $a_0$, in the next section. In the regime $a\gtrsim a_{t}$, where Einstein gravity holds, we have $f(r)=1$. In this regime, $A(r)$ and $B(r)$ are given by the following relations~\cite{weinberg2013gravitation}
\begin{align}
A(r) & =1-\frac{2GM}{r} \, .
\label{A-Schwarzschild}
\\
B(r) & =\left(1-\frac{2GM}{r}\right)^{-1} \, .
\label{B-Schwarzschild}
\end{align}
Substituting these functions into Eq.~\eqref{a-f}, we reach
\begin{equation}
|a|=\frac{GM}{r^{2}}-\frac{2G^{2}M^{2}}{r^{3}}\approx\frac{GM}{r^{2}} \, .
\end{equation}
Here we note that the term proportional to $r^{-3}$ in the above equation is negligible compared to the term proportional to $r^{-2}$ at sufficiently large distances. Using this result in the definition~\eqref{mueff}, we will have
\begin{equation}
\mu_{\mathrm{eff}}\left(|a|\right)=1 \, .
\label{muff1}
\end{equation}
This implies that in the regime $a\gtrsim a_{t}$, the function $\mu_{\mathrm{eff}}$ satisfies the condition given in Eq.~\eqref{mu} for the MOND interpolation function.

Now we consider the case of accelerations smaller than the transition acceleration ($a\lesssim a_{t}$). In this regime, we enter the regime of modified gravity, in which the function $f(r)$ is given by Eq.~\eqref{fapprox} and the gravitational acceleration is given by Eq.~\eqref{absa}. Therefore, in this case, the function $\mu_{\mathrm{eff}}$ is resulted in from Eq.~\eqref{mueff} as
\begin{equation}
\mu_{\mathrm{eff}}=\frac{2GMr^{-\epsilon-1}}{|C_{A}\epsilon|} \, .
\label{mueff-r}
\end{equation}
Now we invert Eq.~\eqref{absa} to write the radial coordinate $r$ in terms of the absolute value of acceleration as
\begin{equation}
r=\left(\frac{2|a|}{|C_{A}\epsilon|}\right)^{\frac{1}{-1+\epsilon}} \, .
\label{r-absa}
\end{equation}
By substituting this into Eq.~\eqref{mueff-r}, we find
\begin{equation}
\mu_{\mathrm{eff}}\left(|a|\right)=4^{\frac{1}{1-\epsilon}}GM|C_{A}\epsilon|^{\frac{2}{-1+\epsilon}}|a|^{\frac{1+\epsilon}{1-\epsilon}} \, .
\label{mueff-absa}
\end{equation}
Given that the parameter $\epsilon$ is much smaller than unity, the above relation can be approximated as
\begin{equation}
\mu_{\mathrm{eff}}\left(|a|\right)\approx\frac{4GM}{|C_{A}\epsilon|^{2}}|a|=\frac{|a|}{a_{0}} \, .
\label{mueff2}
\end{equation}
In this relation, it is clear that in the limit of very weak gravitational fields, the function $\mu_{\mathrm{eff}}$ is almost linearly proportional to $|a|$, and as a result, the necessary condition given in Eq.~\eqref{mu} for the MOND interpolation function is also well satisfied in this regime. In addition, as we saw in the above equation, our model can provide a suggestion relation for the Milgrom acceleration $a_0$. It is expected that the transition acceleration in our setup will be of the order of the Milgrom acceleration.

Given that both conditions given in Eq.~\eqref{mu} are satisfied in the relevant gravitational regimes, we conclude that the phenomenological formula of MOND can be recovered in our modified gravity framework.


\section{Comparison with Observational Data}
\label{sec:obs}

Our main goal in this section is to test the consistency of our relativistic MOND model in light of the observational data. To do so, we evaluate the rotational velocity curves of the stars at the edge of a galaxy in this model and compare the results with the measured values. In our work, we use data from the galaxy NGC 3198, which belongs to the SPARC catalog and whose data are publicly available\footnote{\url{https://astroweb.cwru.edu/SPARC/Rotmod_LTG.zip}}. The catalog we use contains results for 43 stars in which the rotational velocities are listed as a function of distance. In this section, in order to be able to estimate the observational quantities in our model correctly and in the correct dimensions, we need to reintroduce the fundamental constants $G$, $c$, and $\hbar$ to the theoretical equations.

We also compare our RMOND model results with those of the classical Newtonian dynamics (ND) model and the model including dark matter (DM). In the ND and DM models, the gravitational acceleration is given by the following equation
\begin{equation}
a=\frac{GM(r)}{r^{2}} \, .
\label{aND}
\end{equation}
In the ND model, the mass $M(r)$ only includes the baryonic mass of the galaxy disk $M_d(r)$, but in the DM model, this mass of includes both the baryonic mass of the galaxy disk of the galaxy $M_d(r)$ and the mass of the dark matter halo $M_{DM}(r)$. The mass of the galaxy disk is the sum of the mass of the stars in that galaxy, and here we have neglected the mass of the interstellar gas. We assume an exponential profile for the surface mass density of the disk with the radial distance $r$ as~\cite{Freeman:1970mx}
\begin{equation}
\sigma(r)=\sigma_{0}e^{-r/R_{d}} \, .
\label{sigma}
\end{equation}
In this equation, $\sigma_{0}$ is the surface mass density at the center of the galaxy and $R_d$ is the characteristic radius of the disk at which the disk mass density declines to the value of $\sigma_{0}/e$. By integrating the above equation on the surface of the disk, we obtain the mass of the galaxy disk as
\begin{equation}
M_{d}(r)=\int_{0}^{r}(2\pi r')\sigma(r')\,dr'=2\pi R_{d}\sigma_{0}\left[R_{d}-e^{-r/R_{d}}\left(r+R_{d}\right)\right] \, .
\label{Md}
\end{equation}

To calculate the mass of dark matter, it is usually assumed that the mass distribution of the dark matter halo per unit volume follows the following profile~\cite{Begeman:1991iy}
\begin{equation}
\rho_{\mathrm{DM}}(r)=\frac{\rho_{\mathrm{DM0}}}{\left(\frac{r}{r_{c}}\right)^{2}+1} \, ,
\label{rhoDM}
\end{equation}
where $\rho_{\mathrm{DM0}}$ is the density of the dark matter halo at the center of the galaxy and $r_c$ is a critical radius at which the halo density is reduced to half its value at the center. Although most studies interpret this profile as representing the dark matter mass distribution in the galactic halo, there is no clear physical justification for adopting such a profile. By integrating the above equation, the mass of the dark matter halo is obtained as
\begin{equation}
M_{\mathrm{DM}}=\int_{0}^{r}\left(4\pi r'^{2}\right)\rho_{\mathrm{DM}}(r')\,dr'=4\pi\rho_{\mathrm{DM0}}r_{c}^{2}\left[r-r_{c}\arctan\left(\frac{r}{r_{c}}\right)\right] \, .
\label{MDM}
\end{equation}

In the ND model, the rotational velocity relationship is given by the following equation
\begin{equation}
v(r)=\sqrt{\frac{GM_{d}(r)}{r}} \, .
\label{vND}
\end{equation}
In the DM model, the mass of the dark matter halo is added to the mass of the galaxy disk in the above equation, and as a result, we will have
\begin{equation}
v(r)=\sqrt{\frac{G\left[M_{d}(r)+M_{\mathrm{DM}}(r)\right]}{r}} \, .
\label{vDM}
\end{equation}

In order to examine rotational velocities in the RMOND model, it is necessary to introduce the concept of the transition radius $r_t$ at which the gravitational acceleration is equal to the transition acceleration $a_t$. In fact, we consider the transition acceleration $a_t$ as an acceleration at which the gravitational theory has completely changed and entered a very weak gravitational regime in which the RMOND theory effectively holds. It is expected that the transition acceleration $a_t$ is of the order of the reference acceleration $a_\mathrm{ref}$ or slightly less, because $a_\mathrm{ref}$ in fact introduces the threshold value of the acceleration at which the gravitational regime changes commence. In fact, the reference acceleration $a_\mathrm{ref}$, according to its definition, is equivalent to the Milgrom acceleration $a_0$, which is of the order of $10^{-10}\,m/s^{2}$. With these considerations, we expect the acceleration $a_t$ to be of the order of the Milgrom acceleration $a_0$ or slightly less.

In the RMOND model, as in the ND model, it is assumed that the mass of the galaxy only includes the mass of the disk, which is given by Eq.~\eqref{Md}. Also, in this model, it is assumed that for distances less than the transition distance $r_t$, the Newtonian gravitational regime prevails, and as a result, the relationship of the rotational velocity is given by Eq.~\eqref{vND}. However, for distances greater than $r_t$, the gravitational regime changes, and as a result, we use Eq.~\eqref{vRMOND-a} to calculate the rotational velocity. Therefore, in general, to calculate the rotational velocity in the RMOND model, we use the following equation
\begin{equation}
v(r) =
\begin{cases}
\sqrt{\dfrac{G\,M_{d}(r)}{r}}\,, & r \lesssim r_{t},\\[6pt]
\sqrt{\dfrac{|C_{A}\,\epsilon|}{2}}\; c^{\frac{3\epsilon}{4}+1} \,(G\hbar)^{-\frac{\epsilon}{4}} \, r^{\frac{\epsilon}{2}}\,, & r \gtrsim r_{t}.
\end{cases}
\label{vRMOND}
\end{equation}

Now we want to constrain the mentioned models using observational data for the galaxy NGC~3198. For this purpose, we develop a computational code that is publicly available\footnote{\url{https://github.com/krezazadeh/MCMC-RMOND}}. In this code, the multi-chain Monte Carlo (MCMC) method is implemented using the Metropolis-Hastings algorithm. The code is written based on the message passing interface (MPI) method, which executes multiple MCMC chains simultaneously. The chains are periodically synchronized until the Gelman-Rubin diagnostic ($\hat{R}$) is monitored. In our work, we consider the convergence criterion as $\hat{R}-1<0.01$, which guarantees that the MCMC analysis converges well. For the statistical analysis of MCMC chains, we use the GetDist package~\cite{Lewis:2019xzd}, which is publicly available\footnote{\url{https://getdist.readthedocs.io/en/latest/}}.

\begin{figure}
\centering
\includegraphics[width=0.6\textwidth]{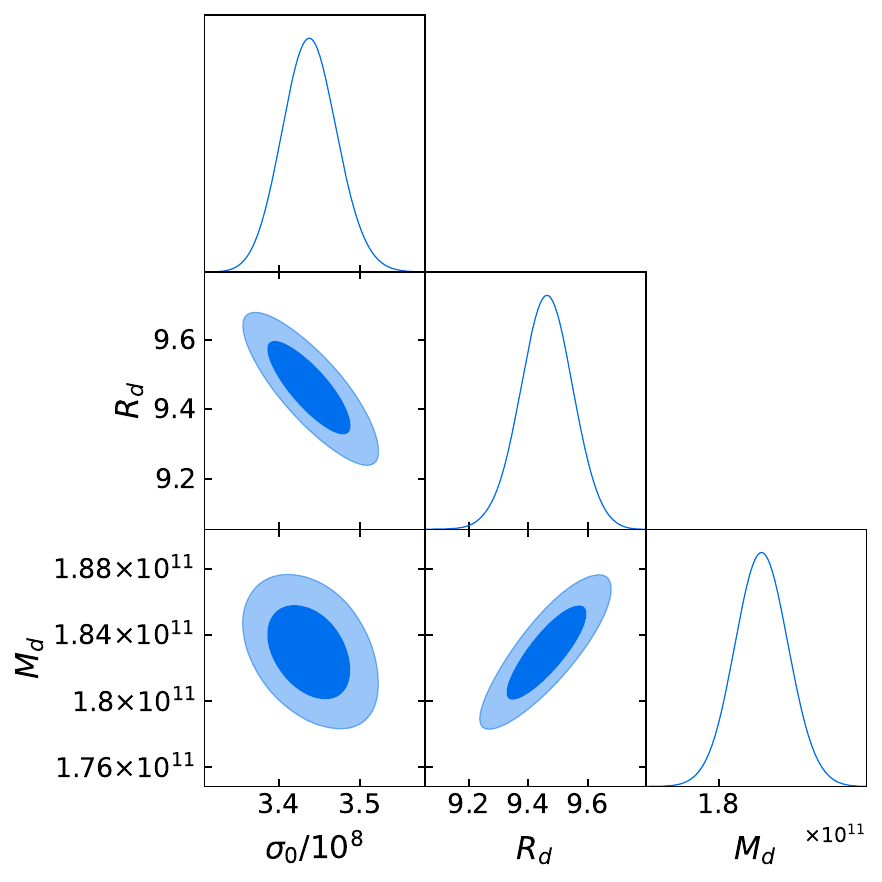}
\caption{1D likelihoods and 2D contours for the parameters in 68\% and 95\% CL marginalized joint regions for the ND model.}
\label{fig:triangle-ND}
\end{figure}

\begin{figure}
\centering
\includegraphics[width=\textwidth]{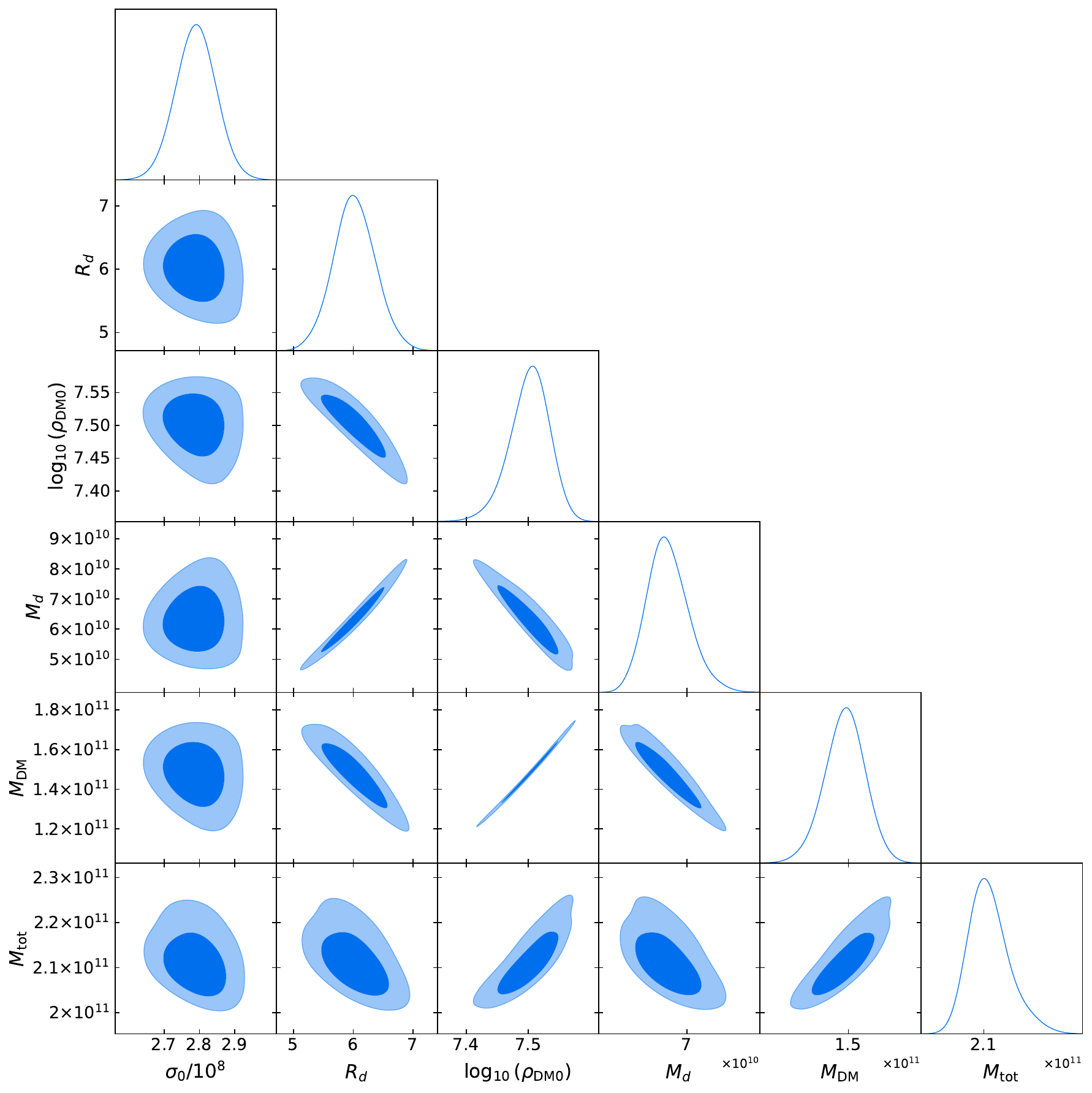}
\caption{1D likelihoods and 2D contours for the parameters in 68\% and 95\% CL marginalized joint regions for the DM model.}
\label{fig:triangle-DM}
\end{figure}

\begin{figure}
\centering
\includegraphics[width=\textwidth]{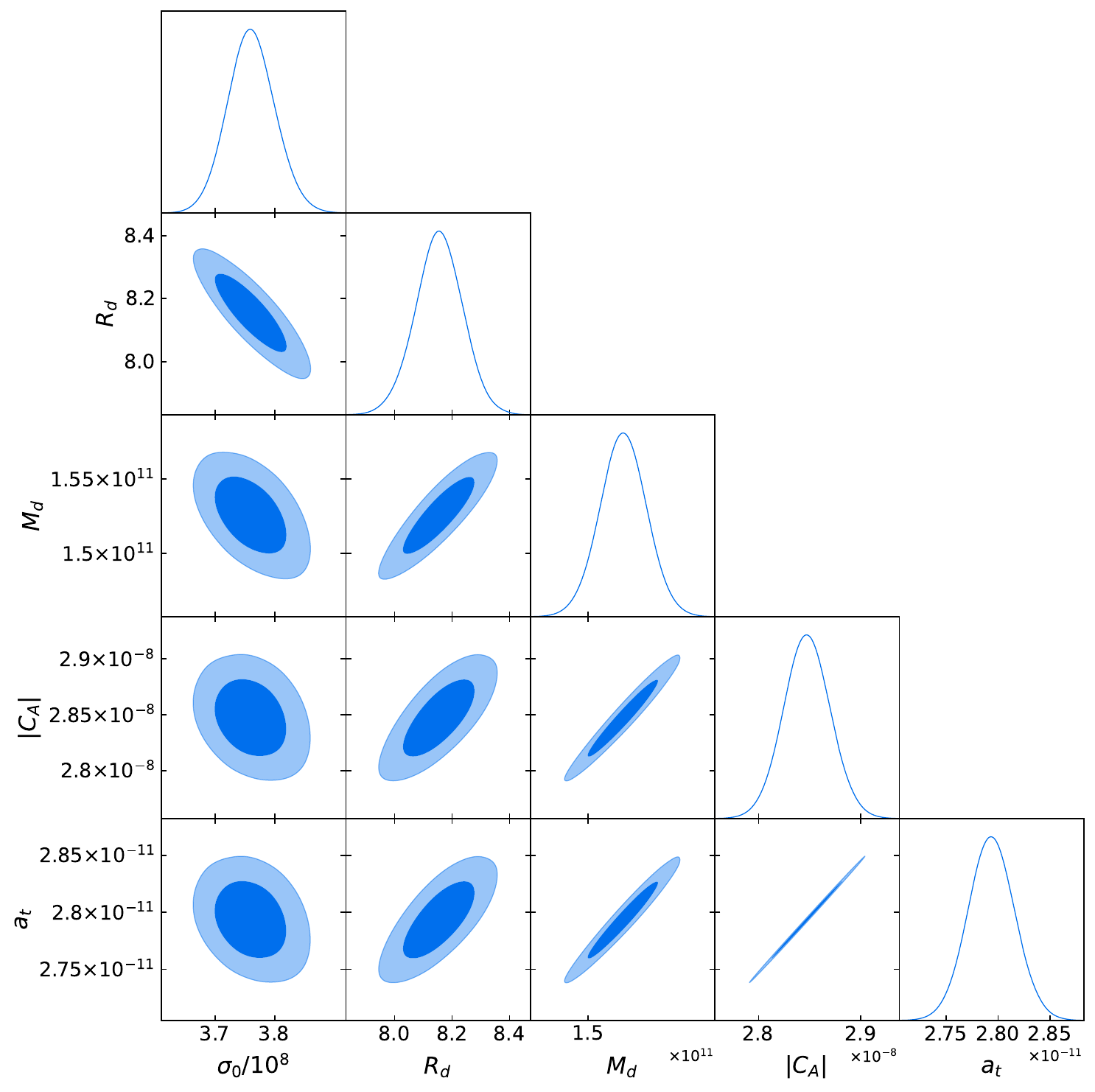}
\caption{1D likelihoods and 2D contours for the parameters in 68\% and 95\% CL marginalized joint regions for the RMOND model.}
\label{fig:triangle-RMOND}
\end{figure}

The results of our MCMC implementation for the best fit value and constraints with a confidence limit (CL) of 68\% for the free and derived parameters in the models under consideration are given in Table~\ref{table:parameters}. In Figs.~\ref{fig:triangle-ND}-\ref{fig:triangle-RMOND}, the triangular plots of the one-dimensional and two-dimensional posteriors of the free and derived parameters resulting from the MCMC analysis for the three models under consideration are shown. In the MD model, a degeneracy is observed between the two free parameters $\rho_{\mathrm{DM0}}$ and $r_c$, and therefore, we have fixed the critical radius to the value $r_{c}=3.062\,\mathrm{kpc}$. In the case of the RMOND model, we also observed that the MCMC method cannot constrain the free parameters $\epsilon$ and $r_t$ properly, and therefore we have fixed these parameters with the values $0.046$ and $25.285\,\mathrm{kpc}$, respectively.

\begin{table*}[!ht]
\centering
\caption{The best-fit values and 68\% CL constraints for the parameters of the investigated models.}
\scalebox{0.8}{
\begin{tabular}{|c|c|c|c|c|c|c|}
\hline
\multirow{2}{*}{Parameter} & \multicolumn{2}{c|}{ND} & \multicolumn{2}{c|}{DM} & \multicolumn{2}{c|}{RMOND}\tabularnewline
\cline{2-7} \cline{3-7} \cline{4-7} \cline{5-7} \cline{6-7} \cline{7-7}
 & best-fit & 68\% limits & best-fit & 68\% limits & best-fit & 68\% limits\tabularnewline
\hline
\hline
$\tilde{\sigma}_{0}/10^{8}$ & $3.467$ & $3.438\pm0.034$ & $2.825$ & $2.790\pm0.058$ & $3.790$ & $3.761\pm0.040$\tabularnewline
\hline
$R_{d}\,[\mathrm{kpc}]$ & $9.376$ & $9.461\pm0.090$ & $5.110$ & $6.02\pm0.36$ & $8.087$ & $8.154\pm0.083$\tabularnewline
\hline
$\log_{10}\left(\tilde{\rho}_{\mathrm{DM0}}\right)$ & $-$ & $-$ & $7.569$ & $7.502_{-0.028}^{+0.035}$ & $-$ & $-$\tabularnewline
\hline
$M_{d}\,[M_{\varodot}]$ & $1.816\times10^{11}$ & $\left(1.830\pm0.020\right)\times10^{11}$ & $4.626\times10^{10}$ & $\left(6.33_{-0.79}^{+0.64}\right)\times10^{10}$ & $1.514\times10^{11}$ & $\left(1.526\pm0.017\right)\times10^{11}$\tabularnewline
\hline
$M_{\mathrm{DM}}\,[M_{\varodot}]$ & $-$ & $-$ & $1.723\times10^{11}$ & $\left(1.48\pm0.11\right)\times10^{11}$ & $-$ & $-$\tabularnewline
\hline
$M_{\mathrm{tot}}\,[M_{\varodot}]$ & $-$ & $-$ & $2.185\times10^{11}$ & $\left(2.114_{-0.056}^{+0.037}\right)\times10^{11}$ & $-$ & $-$\tabularnewline
\hline
$\left|C_{A}\right|$ & $-$ & $-$ & $-$ & $-$ & $2.835\times10^{-8}$ & $\left(2.847\pm0.023\right)\times10^{-8}$\tabularnewline
\hline
$a_{t}\,[m/s^{2}]$ & $-$ & $-$ & $-$ & $-$ & $2.782\times10^{-11}$ & $\left(2.794\pm0.023\right)\times10^{-11}$\tabularnewline
\hline
\end{tabular}
}
\label{table:parameters}
\end{table*}

\begin{table*}[!ht]
\centering
\caption{The resulting values of $\chi^2$ for each model and each data set. Here, we have defined $\Delta\chi^{2}=\chi_{\mathrm{model}}^{2}-\chi_{\mathrm{DM}}^{2}$.}
\scalebox{1.0}{
\begin{tabular}{|c|c|c|c|}
\hline
$\qquad$ $\qquad$ & $\qquad$ ND $\qquad$ & $\qquad$ DM $\qquad$ & $\qquad$ RMOND $\qquad$ \tabularnewline
\hline
\hline
$\chi^{2}$ & $277.26$ & $57.12$ & $61.37$\tabularnewline
\hline
$\Delta\chi^{2}$ & $220.14$ & $0.0$ & $4.25$\tabularnewline
\hline
\end{tabular}
}
\label{table:chi2}
\end{table*}

We see in Table~\ref{table:parameters} that in the ND and RMOND models, the mass of the galaxy disk $M_d$ is of the order $10^{11}\,M_{\astrosun}$, but in the DM model, it is of order $10^{11}\,M_{\astrosun}$ which is one order of magnitude smaller. These predictions can be used to distinguish between these models by considering other methods for measuring the mass of the galactic disk. We further see in the table that in our model, the transition acceleration $a_t$ is obtained of the order $10^{-11}\,m/s^{2}$, which is one order of magnitude less than the value for the Milgrom acceleration $a_{0}\sim10^{-10}\,m/s^{2}$. Of course, here we have calculated $a_t$ at a moment when it is assumed that the new gravitational regime is fully established, so the acceleration at this time is expected to be less than the Milgrom acceleration $a_0$ that determines the threshold for changing the gravitational regime.

\begin{figure}
\centering
\includegraphics[width=0.6\textwidth]{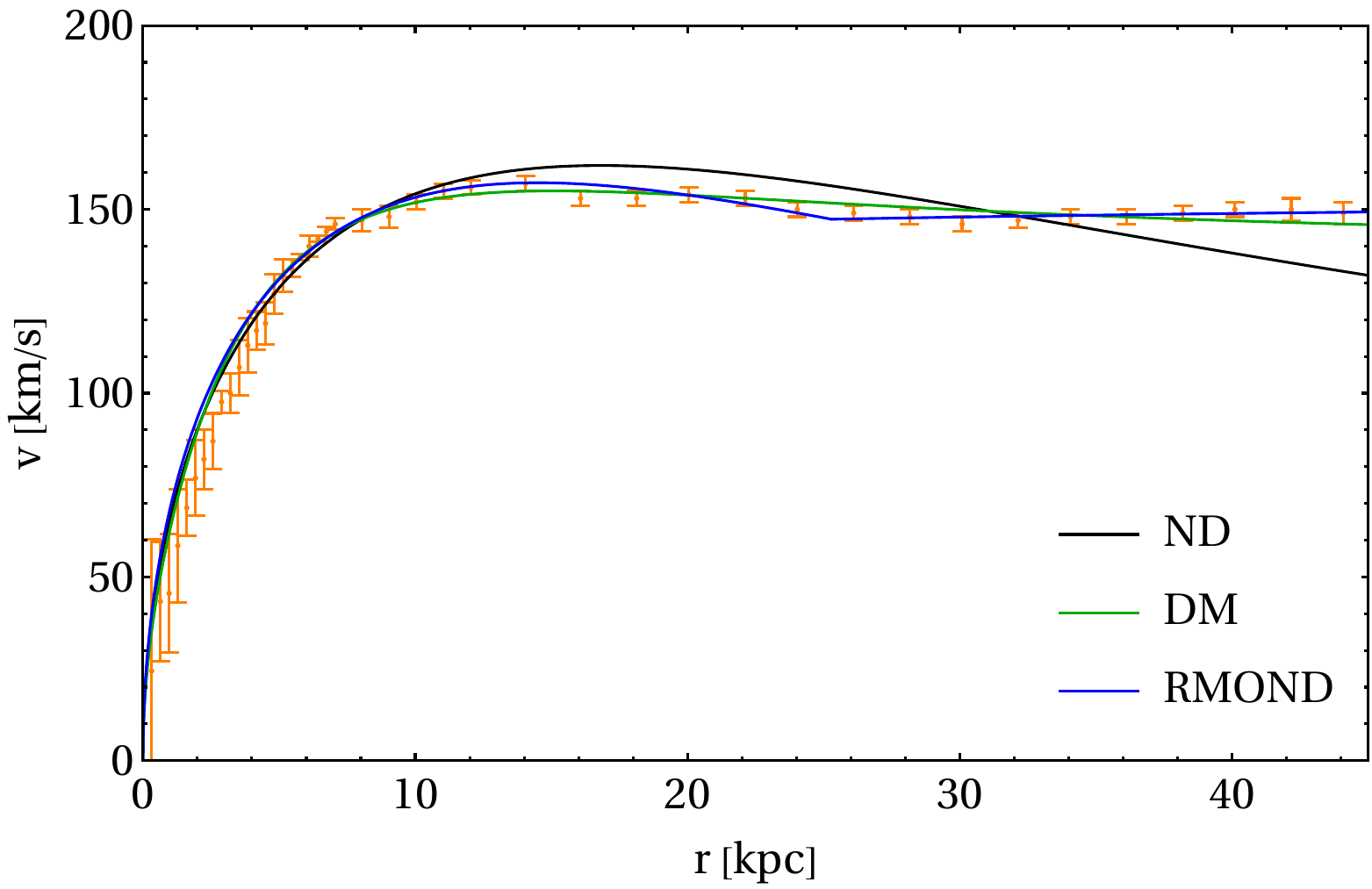}
\caption{Rotational velocity versus distance from the galaxy center for the ND, DM, and RMOND models. The data points from the measurements of the galaxy NGC~3198 are also presented in the figure.}
\label{fig:v-r}
\end{figure}

In Table~\ref{table:chi2}, the results of the values of the $\chi^2$ quantity in the MCMC implementation for the models under consideration are presented. The following relation gives this quantity as
\begin{equation}
\chi^{2}=\underset{i}{\sum}\frac{\left(v^{(\mathrm{th})}(r_{i})-v^{(\mathrm{obs})}(r_{i})\right)}{\sigma_{i}^{2}} \, ,
\label{chi2}
\end{equation}
where $v^{(\mathrm{th})}(r_{i})$ is equal to the value obtained from the theoretical model under consideration for the rotational velocity at the radial distance $r_i$, while $v^{(\mathrm{obs})}(r_{i})$ denotes the measured value for it. In addition, $\sigma_{i}$ represents the error in the measurement of the rotational velocity. The lower the value of $\chi^{2}$ a theoretical model gives, the better the fit of that model to the observational data. We see in Table~\ref{table:chi2} that the value of $\chi^2$ in the DM and RMOND models is much lower than the value obtained in the ND model. Therefore, the two models, MD and RMOND, in contrast to the ND model, provide a good fit to the observational data. The DM model gives the lowest value for the $\chi^2$ parameter and, as a result, the best fit is associated with this model. But as we said before, this mass uses a mass profile that has no physical justification. However, our analysis indicates that for large radial distances, namely for $r\gtrsim20\,\mathrm{kpc}$, the RMOND scenario provides a considerably better fit to the observational data. To declare this point more clearly, we have plotted the variation of the rotational velocity with distance in Fig.~\ref{fig:v-r}, for the models under consideration. By comparing the plots of the MD and RMOND model in this figure, it is evident that the RMOND model fits the data much better than the DM model in the radial distances of $r\gtrsim20\,\mathrm{kpc}$.

\section{Conclusions}
\label{sec:con}

In this work, we have explored a relativistic extension of Modified Newtonian Dynamics (MOND) within the framework of entropic gravity. By incorporating temperature-dependent corrections to the equipartition law, one can derive modified Einstein equations that introduce explicit thermal corrections to the geometric sector of gravity. These modifications naturally lead to a characteristic acceleration scale, $a_0$, which is consistent with the phenomenology of MOND. This approach provides a theoretical underpinning for MOND, grounded in thermodynamic principles, and extends the framework of entropic gravity proposed by Verlinde~\cite{Verlinde:2010hp}.

The solutions to the modified Einstein equations for a static, spherically symmetric spacetime showed that the metric differs from the standard Schwarzschild solution, yet smoothly reduces to Minkowski space in the low-temperature limit. For accelerations less than a characteristic acceleration, the metric form of the Schwarzschild relation changes to a modified form that determines the geometry of space-time in the very weak gravity limit. This result offers a relativistic foundation for MOND, which has traditionally been a non-relativistic theory.

From an astrophysical perspective, we applied the relativistic MOND (RMOND) model to the galaxy NGC~3198, comparing its predictions with those of classical Newtonian dynamics (ND) and the dark matter (DM) model. The Bayesian parameter inference showed that the RMOND model provides a better fit to the data at large galactocentric radii ($r \gtrsim 20\,\mathrm{kpc}$), compared to both the Newtonian and dark matter models. This finding is in line with earlier work, such as~\cite{Milgrom:1983ca} and \cite{Famaey:2011kh}, which also explored the validity of MOND in explaining galactic dynamics without invoking dark matter. The RMOND model achieved strong consistency with observational data, providing an alternative explanation to dark matter for galactic rotation curves, especially at large distances where dark matter halos become increasingly difficult to constrain.

Notably, the RMOND model offers a compelling explanation for the observed rotational velocity curves of galaxies, producing a smooth transition from Newtonian dynamics to a MOND-like regime at large radii. This stands in contrast to the dark matter model, which typically requires a halo profile with a specific mass distribution that lacks clear physical justification. The RMOND model, in contrast, requires only the mass of the galaxy disk, offering a simpler, non-dark-matter-based explanation for the observed data. 

Although our relativistic extension of MOND successfully accounts for galactic rotation curves without dark matter in the weak-field regime, it does not, in its current form, address cluster-scale dynamics, gravitational lensing, or systems such as the Bullet Cluster. These phenomena involve additional relativistic, lensing, and cosmological effects that lie outside the focus of this work. A comprehensive treatment of these observational tests would require extending the present framework and is left for future research.

\subsection{Discussion of Results}

The results of our analysis highlight the potential of entropic gravity as a framework for understanding modified gravity at galactic scales. The introduction of temperature-dependent corrections to the equipartition law leads to a modified spacetime geometry that naturally accommodates the acceleration scale $a_0$ seen in MOND-like deviations from Newtonian gravity. Our analysis also demonstrated that, unlike the standard MOND framework, the RMOND model incorporates relativistic effects, making it more robust for cosmological and large-scale applications.

The comparison with observational data for NGC~3198 showed that RMOND provides an excellent fit to rotation curves, especially at large radii where the dark matter model struggles to capture the true dynamics. In contrast, the purely Newtonian model fails to explain the flatness of rotation curves, a signature feature of MOND. The results suggest that temperature-corrected entropic gravity can successfully reproduce the key features of MOND at galactic scales, while offering a coherent and relativistic framework that avoids the need for dark matter.

\subsection{Future Directions}

While the RMOND model provides a promising framework for understanding galactic dynamics, several avenues remain for further investigation:

\begin{itemize}

\item Gravitational Lensing: One of the most promising directions is to extend the RMOND framework to gravitational lensing. Given that lensing is a direct probe of spacetime geometry, comparing the predictions of RMOND with observed lensing data could provide a critical test of the model. Future observations from the Euclid mission~\cite{Euclid:2024yrr} or the Vera C. Rubin Observatory~\cite{LSST:2008ijt} could offer significant constraints.

\item Cosmological Evolution: The implications of RMOND for the large-scale evolution of the universe should be explored. In particular, the role of entropic gravity and thermal corrections in cosmic inflation, the formation of large-scale structures, and the behavior of the cosmic microwave background (CMB) deserves further attention. Investigating the effects of RMOND on cosmological parameters would help assess its broader implications for the standard cosmological model.

\item Galaxy Clusters and Larger Scales: Applying the RMOND model to galaxy clusters, where the dynamics are more complex, would provide additional tests of its validity at larger scales. This could help clarify whether RMOND can successfully explain the behavior of galaxy clusters, where dark matter models typically dominate.

\item Quantum Gravity: Another exciting avenue is to explore the connections between entropic gravity and quantum gravity. The emergence of gravity from statistical mechanics, as suggested by the holographic principle and the Unruh effect, could provide new insights into the fundamental nature of gravity at the quantum level. Investigating these connections could lead to a deeper understanding of the potential unification of gravity with quantum mechanics.

\item Multi-Galaxy Datasets: Finally, a broader observational study using multi-galaxy datasets from upcoming astronomical surveys, such as those planned for the Vera C. Rubin Observatory~\cite{LSST:2008ijt}, would provide a more comprehensive test of the model’s ability to explain galactic dynamics across different galaxy types, redshifts, and environments.

\end{itemize}

\subsection{Final Remarks}

In summary, this work successfully constructs a relativistic extension of MOND, grounded in entropic gravity, that provides a natural and dark-matter-free explanation for galactic rotation curves. The RMOND model achieves strong agreement with observational data, particularly at large galactic radii, where dark matter models are less constrained. As we continue to refine and extend this framework, the potential for RMOND to reshape our understanding of gravity and cosmology becomes increasingly apparent. With further observational and theoretical work, RMOND may offer a compelling alternative to the dark matter paradigm and provide a deeper understanding of the nature of gravity in our universe.


\section*{Acknowledgements}

The authors are grateful to the referee for their valuable and insightful comments, which have helped improve the clarity and quality of the manuscript.


%
%


\bibliography{RMOND}


\end{document}